\documentclass[prl,aps,showpacs,twocolumn,floatfix]{revtex4}
\usepackage{epsfig} \usepackage{graphics} \usepackage{bm}
\usepackage{amssymb}
\begin{document}

\preprint{Lebed-Wu-PRL}

\title{Larkin-Ovchinnikov-Fulde-Ferrell Phase in (TMTSF)$_2$ClO$_4$ Superconductor:
Theory versus Experiment}

\author{A.G. Lebed$^*$}
\author{Si Wu}

\affiliation{Department of Physics, University of Arizona, 1118 E.
4-th Street, Tucson, AZ 85721, USA}

\begin{abstract}
We consider a formation of the Larkin-Ovchinnikov-Fulde-Ferrell (LOFF) phase
in a quasi-one-dimensional (Q1D) conductor in a magnetic field, parallel to its
conducting chains, where we take into account both the paramagnetic spin-splitting 
and orbital destructive effects against 
superconductivity.
We show that, due to a relative weakness of the orbital effects in a Q1D case,
the LOFF phase appears in (TMTSF)$_2$ClO$_4$ superconductor for real values of
its Q1D band parameters.
We compare our theoretical calculations with the recent experimental data by
Y. Maeno's group [S. Yonezawa et al., Phys. Rev. Lett. \textbf{100}, 117002 (2008)]
and show that there is a good qualitative and quantitative agreement between the 
theory and experimental data.
\end{abstract}

\pacs{74.70.Kn, 74.20.Rp, 74.25.Op}

\maketitle


Since a discovery of superconductivity in organic (TMTSF)$_2$X conductors
(X=PF$_6$ and ClO$_4$) [1], their physical properties have been intensively
studied both experimentally and theoretically [2,3].
From the beginning, it was clear that their superconducting properties were unconventional.
Indeed, it was found [4] that superconductivity was destroyed by non-magnetic impurities, which was recently unequivocally 
 confirmed [5].
In addition, it was shown [6] that the conventional for s-wave superconductivity
Hebel-Slichter peak was absent in the NMR experiments.
Note that the experimental results [4-6] provide strong arguments that superconducting
order parameter changes its sign on a quasi-one-dimensional (Q1D) Fermi surface
(FS) of (TMTSF)$_2$X superconductors.
On the other hand, they do not contain information about a spin-part of a superconducting order parameter and, thus, do not distinguish between
singlet and triplet pairings.
At the moment, the problem about a spin part of a superconducting order parameter
in (TMTSF)$_2$X conductors is still controversial.
Indeed, early measurements of the Knight shift in (TMTSF)$_2$PF$_6$ conductor
[7] in a magnetic field $H = 1.43 \ T$ showed that spin susceptibility was unchanged
through the superconducting transition.
These data were interpreted as an evidence for triplet superconductivity.
On the other hand, more recent Knight shift data in (TMTSF)$_2$ClO$_4$
conductor [8], obtained in a magnetic field $H = 0.957 \ T$, were interpreted in favor 
of a singlet
superconducting pairing.

Another unconventional feature of superconductivity in (TMTSF)$_2$X
conductors is very large upper critical fields for a magnetic field parallel to
their conducting planes and perpendicular to their conducting chains,
${\bf H} \parallel {\bf b'}$ [9-14].
These fields exceed both the quasi-classical orbital upper critical field [15,16] and 
so-called Clogston paramagnetic limit [17].
Note that in (TMTSF)$_2$PF$_6$ superconductor $H^{b'}_{c2}$ is very large [10,11]
due to a formation of domain walls in the vicinity of antiferromagnetic phase.
In contrast, in (TMTSF)$_2$ClO$_4$ conductor large upper critical field $H^{b'}_{c2}$
[12-14] is prescribed to $3D \rightarrow 2D$ dimensional crossover in a magnetic field, predicted in Ref.[18] and elaborated in 
Refs. [19-23].
In addition, recent measurements of the upper critical fields in (TMTSF)$_2$ClO$_4$
superconductor  [13,14] have revealed another unusual property - a novel superconducting phase, which appears when a magnetic field is applied along 
the conducting
chains, ${\bf H} \parallel {\bf a}$.
It is shown [13,14] that the above mentioned phase is very sensitive to impurities
and inclinations of a magnetic field from 
${\bf a}$ axis.
The authors of the experiments [13,14] have related this new phase with a possible
formation of the Larkin-Ovchinnikov-Fulde-Ferrell (LOFF) state [24,25].

The goal of our Letter is to show theoretically that the LOFF phase has to appear
in a magnetic field, parallel to the conducting chains of (TMTSF)$_2$ClO$_4$
superconductor, if we use experimentally measured values of its Q1D band
parameters.
The distinctive feature of our work is that we take into account both the paramagnetic
spin-splitting and orbital destructive effects against
superconductivity.
This problem is a challenging one, since the orbital effects
for a magnetic field along the conducting chains correspond to very particular
open electron trajectories.
These effects cannot be described by any of the existing theories, including
Refs. [15-23,26].
To describe the orbital and paramagnetic effects, below we derive an integral 
equation for a superconducting order parameter, which, to the best of our knowledge, 
has not been considered 
before.
By means of the measured Ginzburg-Landau slops of an anisotropic upper
critical field in (TMTSF)$_2$ClO$_4$ [13,14] and the measured ratio $t_a/t_b$
[27,28], we determine its Q1D band
parameters.
We use these band parameters for a numerical solution of the above mentioned integral
equation and conclude that the LOFF phase has to exist in (TMTSF)$_2$ClO$_4$
conductor.
This conclusion is based both on theoretical arguments and on good qualitative
and quantitative agreement between the theory and experimental
measurements [13,14].
It is important that the results of our Letter support singlet d-wave 
scenario of superconductivity in (TMTSF)$_2$X
materials.

Below, we consider a Q1D conductor with the following electron spectrum,
\begin{equation}
\epsilon({\bf p})= - 2t_a \cos(p_x a/2) - 2 t_b \cos(p_yb^*) - 2t_c
\cos (p_zc^*),
\end{equation}
in a magnetic field, parallel to its conducting chains, ${\bf
H}\parallel {\bf a}$,
\begin{equation}
{\bf H} = (H,0,0), \ \ {\bf A} = (0,0,Hy),
\end{equation}
where $t_a \gg t_b \gg t_c$ correspond to electron hoping integrals along
${\bf a}$ , ${\bf b}$, and ${\bf c}$ axes, respectively. 

We represent electron wave functions with definite energy and momentum
$p_x$ in the following way:
\begin{equation}
\Psi_{\epsilon}^{\pm}(p_x;x,y) = \exp(\pm ip_x x) \exp[\pm ip^{\pm}_y(p_x)y] \ 
\psi_{\epsilon}^{\pm}(p_x,y) ,
\end{equation}
where +(-) stands for left (right) sheet of a Q1D FS and the functions $p_y^{\pm}(p_x)$ are defined by the equations:
\begin{equation}
v_F(p_x \mp p_F) \mp 2 t_b \cos[p^{\pm}_y(p_x)b^*] =0,
\end{equation}
where $v_F$ and $p_F$ are the Fermi velocity and Fermi momentum,
respectively.
In this case, we can rewrite Eq. (1) as:
\begin{equation}
\delta \epsilon^{\pm}({\bf p})= \pm 2 t_b b^* (p_y-p_y^{\pm}) \sin(p^{\pm}_yb^*) -
2t_c \cos (p_zc^*) ,
\end{equation}
where we linearize the electron spectrum (1) near the FS with
respect to momentum $p_y$; energy $\delta \epsilon = \epsilon -  \epsilon_F$
is counted from the Fermi energy $\epsilon_F$.

In a magnetic field, we use the Peierls substitution method [18] for
Eq.(5),
\begin{equation}
p_y-p^{\pm}_y \rightarrow -i\frac{d}{dy}, \ p_zc^* \rightarrow
p_zc^* - \frac{\omega_c}{v_F}y, \ \omega_c=\frac{eH v_F c^*}{c},
\end{equation}
where $e$ is the electron charge and $c$ is the velocity 
of light.
As a result, we obtain the following Schrodinger equation for
the electron wave functions:
\begin{eqnarray}
[ \mp i v_y(p^{\pm}_y) \frac{d}{dy} &&- 2t_c \cos (p_zc^*-\frac{\omega_c}{v_F}y)
- \mu_B \sigma H] \ \psi_{\epsilon}^{\pm}(p_x,y,p_z)
\nonumber\\
&&= \delta \epsilon \ \psi_{\epsilon}^{\pm}(p_x,y,p_z),
\end{eqnarray}
with $\sigma/2$ being a projection of an electron spin on ${\bf a}$ axis;
$\mu_B$ is the Bohr magneton,
$v_y(p_y)=2t_bb^* \sin(p_yb^*)$.

It is important that Eq.(7) can be exactly solved:
\begin{eqnarray}
\psi_{\epsilon}^{\pm}(p_x,y,p_z)&&= \frac{1}{\sqrt{2 \pi |v_y(p^{\pm}_y)|}}
\exp  \biggl[ \frac{\pm i \delta \epsilon y}{v_y (p^{\pm}_y)} \biggl]
\exp \biggl[ \frac{\pm i \mu_B \sigma H y}{v_y(p^{\pm}_y)} \biggl]
\nonumber\\
&&\times \exp  \biggl[\pm i \frac{2t_c}{v_y (p^{\pm}_y)} \int^y_0 
\cos \biggl( p_z c^* - \frac{\omega_c}{v_F} u \biggl) du \biggl],
\end{eqnarray}
where the wave functions (8) are normalized by condition:
\begin{equation}
\int^{+\infty}_{-\infty} [\psi^{\pm}_{\epsilon_1}(p_x,y,p_z)]^* \ 
\psi^{\pm}_{\epsilon_2}(p_x,y,p_z) \ dy= \delta (\epsilon_1-\epsilon_2).
\end{equation}
Note that finite temperature Green functions for the wave functions (8),(3)
can be determined by the standard equation:
\begin{equation}
g_{i \omega_n}^{\pm}(x,x_1;y,y_1;p_z)=
\int^{+\infty}_{-\infty} d(\delta \epsilon) \frac{ [\psi^{\pm}_{\epsilon}(x_1,y_1,p_z)]^* \ 
\psi^{\pm}_{\epsilon}(x,y,p_z)}{i \omega_n - \delta \epsilon},
\end{equation}
where $\omega_n$ is the Matsubara frequency [29].

Below, we consider a singlet d-wave scenario of superconductivity
in (TMTSF)$_2$ClO$_4$ conductor, which is in agreement with the
experiments [4-6,8]. 
For this purpose we introduce the following d-wave like LOFF order parameter,
\begin{equation}
\Delta_q(x,y;p_y) =\sqrt{2} \cos(p_yb^*)  \exp(i q x)  \Delta_q(y) ,
\end{equation}
where the Cooper pairs are characterized by non-zero total momentum
along ${\bf a}$ axis, $q \neq 0$, and $\Delta_q(y)$ is the Ginzburg-Landau 
(GL) order parameter, which depends on 
coordinate $y$. 
To derive the gap equation for $\Delta_q(y)$, we use the
Gor'kov equations for non-uniform superconductivity [30,31], as it is
done, for example, in Ref.[32].
As a result of lengthy but rather straightforward calculations, we obtain:

\begin{eqnarray}
&&\Delta_q(y) = g \int \frac{d p_y}{v_x(p_y)} \int_{|y-y_1| > \frac{|v_y|}{\Omega}} 
\frac{2 \pi T dy_1}{v_y(p_y) \sinh \bigl[ \frac{2 \pi T |y-y_1|}{v_y(p_y)} \bigl]}
\nonumber\\
&& \times 2 \cos^2 (p_y b^*) \cos \biggl[ \frac{2 \beta \mu_B H (y-y_1)}{v_y(p_y)} \biggl] \
 \cos \biggl[ q \frac{v_x(p_y)}{v_y(p_y)} (y-y_1) \biggl]
\nonumber\\
&&\times J_0 \biggl\{ 
\frac{8 t_c v_F}{\omega_c v_y(p_y)} 
\sin \biggl[ \frac{\omega_c (y-y_1)}{2v_F} \bigg]
\sin \biggl[ \frac{\omega_c (y+y_1)}{2v_F} \bigg] 
\biggl\} 
\Delta_q (y_1) ,
\end{eqnarray}
where factor $\beta$ takes into account possible decrease of the spin-splitting paramagnetic effects due to small deviations from a weak coupling 
scenario; $g$ is an effective electron coupling 
constant, $\Omega$ is a cutoff energy.
Note that the last term in Eq.(12) describes the orbital effects against
superconducting at low enough magnetic fields.
At high magnetic fields, quantum effects due to the Bragg reflections
of electrons from boundaries of the Brillouin zone, as shown in  Refs. 
[18,32], can improve 
superconductivity.
This improvement may cause to the appearance of the Reentrant
Superconducting (RS) phase [18-23,32].
Our analysis of Eq.(12) shows that the RS phase may appear only
at magnetic fields of the order of $50 \ T$, where superconductivity
is already destroyed by the paramagnetic spin-splitting effects.
Therefore, for a magnetic field parallel to the conducting axis, 
${\bf H} \parallel {\bf a}$, unlike the situation, where it is perpendicular 
to it, ${\bf H} \parallel {\bf b'}$ [18-23], we can ignore the above mentioned
RS effects. 
From mathematical point of view, this means that we can replace the term
$\sin[\frac{\omega_c(y-y_1)}{2v_F}]\sin[\frac {\omega_c(y+y_1)}{2v_F}]$ by
$\frac{\omega^2_c(y^2-y^2_1)}{4v^2_F}$ in Eq.(12).
If we do this and if we introduce more convenient variables, $y=y$, 
$y_1=y+\frac{v_y(p_y)}{v_F}z$, we can rewrite Eq.(12) as:

\begin{eqnarray}
&&\Delta_q(y) = g \int \frac{dp_y}{v_x(p_y)} \int_{|z| > \frac{v_F}{\Omega}} \frac{2 \pi T dz}{v_F \sinh \bigl[ \frac{2 \pi T z}{v_F} \bigl]} 
\nonumber\\
 &&\times 2 \cos^2 (p_y b^*) \cos \biggl[ \frac{2 \beta \mu_B H z}{v_F} \biggl] \
 \cos \biggl[ q \frac{v_x(p_y)}{v_F} z \biggl]
\nonumber\\
&&\times J_0 \biggl\{ \frac{2t_ceHc^*}{v_Fc} z \biggl[2y
+\frac{v_y(p_y)}{v_F}z \biggl] \biggl\} \
\Delta_q \biggl[ y+\frac{v_y(p_y)}{v_F} z \biggl] .
\end{eqnarray}
It is possible to show that in the vicinity of transition temperature,
$(T_c-T) \ll T_c$, Eq.(13) is equivalent to the GL expression for the
upper critical field,
\begin{equation}
H^a_{c2} = \frac{4 \sqrt{2} \pi^2 c \hbar T^2_c}{7 \zeta(3) e t_b t_c b^* c^*}
\biggl( \frac{T_c-T}{T_c} \biggl),
\end{equation}
where $\zeta(x)$ is the Riemann zeta function.

Let us briefly discuss how to determine band parameters $t_a, t_b$, and $t_c$
from analysis of the available experimental data in (TMTSF)$_2$ClO$_4$
conductor [33].
First of all, the ratio $t_a/t_b \simeq 10$ was firmly established in Ref.[28],
where the measured Lee-Naughton-Lebed oscillations were compared with
the corresponding theoretical results.
Second, in Refs.[13,14], the GL slopes for the upper critical fields along
${\bf b}$ and ${\bf c}$ axes were carefully measured: $dH^b_{c2}/dT = 2.75 \ T/K$
and $dH^c_{c2}/dT = 0.14 \ T/K$.
If we take into account that for d-wave like order parameter (11) in the GL 
region [33]:

\begin{equation}
H^b_{c2} = \frac{8  \pi^2 c \hbar T^2_c}{7 \zeta(3) e t_a t_c a c^*}
\biggl( \frac{T_c-T}{T_c} \biggl)
\end{equation}
and
\begin{equation}
H^c_{c2} = \frac{8 \sqrt{2} \pi^2 c \hbar T^2_c}{7 \zeta(3) e t_a t_b a b^*}
\biggl( \frac{T_c-T}{T_c} \biggl),
\end{equation}
then we can evaluate all three band parameters: $t_a = 1340 \ K$, $t_b = 134 \ K$,
$t_c = 2.6 \ K$. 
In Eqs.(15),(16), we use the following known values: $T_c = 1.45 \ K$,
$a=7.1 \ \AA$, $b^*=7.2 \ \AA$, and $c^*=13.1 \ \AA$ [13,14]. 

In Fig.(1), we compare the result of numerical solutions of Eq.(13) for $\beta =0.9$
with the experimental data [13] and predict the appearance of the LOFF phase
at $H \geq 3 \ T$ (see also the figure caption).
Note that the only fitting parameter in our theory is $\beta$. 
We interpret the above mentioned value $\beta =0.9$, which fits better the experimental 
data [13], as an evidence of small deviations from a weak coupling scenario of superconductivity, where 
$\beta=1$.

\begin{figure}[t]
\begin{center}
\epsfig{file=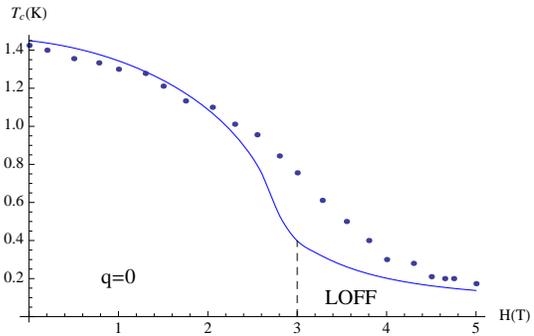,width=7cm} \caption{\label{fig1}
Numerical solution of Eq.(13)  (solid line) is compared to the experimental data 
[13] (dottes).
Note that a phase transition from uniform superconductivity (q=0) to the Larkin-Ovchinnikov-Felde-Ferrell phase ($q \neq 0$) is predicted by our theory
in a magnetic field $H \simeq 3 \ T$.
We pay attention on a good overall qualitative and quantitative agreement
between the theory and experiment.}
\end{center}
\end{figure}

Here, we summarize the main results of the Letter.
We have derived an integral equation, which allows to describe
superconductivity in a magnetic field, parallel to conducting axes of a Q1D
conductor.
By using numerical solutions of the above mentioned equation for experimentally measured band parameters of (TMTSF)$_2$ClO$_4$ compound, we have found 
that the LOFF phase has to appear in this superconductor at high enough magnetic fields.
Comparison of our theoretical curve with the recent experimental data [13] 
demonstrates a good overall qualitative and quantitative 
agreement (see Fig.1).

For theoretical discussions of a possibility of the appearance of the LOFF phase
in a Q1D conductor in a magnetic field, perpendicular to conducting axes and parallel 
to conducting planes, see
Refs.[18,20-23].
There are some experimental data in a favor of the existence of the LOFF phase in quasi-two-dimensional (Q2D) superconductors $\kappa$-(ET)$_2$Cu(NCS)$_2$ [34,35], $\lambda$-(BETS)$_2$GaCl$_4$ [36], and 
$\lambda$-(BETS)$_2$FeCl$_4$ [37].
Nevertheless, theoretical calculations, which take into account both the orbital and
paramagnetic effects in a Q2D case, have not been done yet .
The paramagnetic spin-splitting effects in the LOFF phase in a pure 1D 
case were theoretically studied in Ref. [38], whereas they were theoretically 
studied in a pure 2D case in a number of papers [39-43].

One of us (A.G.L.) is thankful to N.N. Bagmet (Lebed)
for very useful discussions.
This work was supported by the NSF under Grant No DMR-0705986.

$^*$Also at: Landau Institute for Theoretical Physics,
2 Kosygina Street, Moscow, Russia.

\end{document}